\documentclass[pdflatex,sn-mathphys]{sn-jnl}

\jyear{2021}%

\usepackage{amsmath}
\usepackage{amssymb}
\usepackage{mathtools}
\usepackage{booktabs}
\usepackage[binary-units = true, free-standing-units=true]{siunitx}
\DeclareSIUnit\px{px}
\DeclarePairedDelimiter\abs{\lvert}{\rvert}%

\usepackage{soul,xcolor}

\begin{document}

\title[Ensemble uncertainty for upper-body CT images]{Ensemble uncertainty as a criterion for dataset expansion in distinct bone segmentation from upper-body CT images}


\author*[1]{\fnm{Eva} \sur{Schnider}}\email{eva.schnider@unibas.ch}

\author[1]{\fnm{Antal} \sur{Huck}}
\author[2]{\fnm{Mireille} \sur{Toranelli}}
\author[1]{\fnm{Georg} \sur{Rauter}}
\author[1]{\fnm{Azhar} \sur{Zam}}
\author[2]{\fnm{Magdalena} \sur{Müller-Gerbl}}
\author[1]{\fnm{Philippe} \sur{Cattin}}

\affil*[1]{\orgdiv{Department of Biomedical Engineering}, \orgname{University of Basel}, \orgaddress{\street{Gewerbestrasse 14}, \city{Allschwil}, \postcode{4123} \country{Switzerland}}}

\affil[2]{\orgdiv{Department of Biomedicine}, \orgname{University of Basel}, \orgaddress{\city{Basel}, \country{Switzerland}}}

\abstract{
\textbf{Purpose:} 
The localisation and segmentation of individual bones is an important preprocessing step in many planning and navigation applications. 
It is, however, a time-consuming and repetitive task if done manually. 
This is true not only for clinical practice but also for the acquisition of training data. 
We therefore not only present an end-to-end learnt algorithm that is capable of segmenting 125 distinct bones in an upper-body CT, but also provide an ensemble-based uncertainty measure that helps to single out scans to enlarge the training dataset with.

\textbf{Methods:} 
We create fully automated end-to-end learnt segmentations using a neural network architecture inspired by the 3D-Unet and fully supervised training. 
The results are improved using ensembles and inference-time augmentation. 
We examine the relationship of ensemble-uncertainty to an unlabelled scan's prospective usefulness as part of the training dataset.

\textbf{Results:} 
Our methods are evaluated on an in-house dataset of 16 upper-body CT scans with a resolution of \SI{2}{\milli\meter} per dimension. 
Taking into account all 125 bones in our label set, our most successful ensemble achieves a median dice score coefficient of 0.83. 
We find a lack of correlation between a scan's ensemble uncertainty and its prospective influence on the accuracies achieved within an enlarged training set.
At the same time, we show that the ensemble uncertainty correlates to the number of voxels that need manual correction after an initial automated segmentation, thus minimising the time required to finalise a new ground truth segmentation.

\textbf{Conclusion:} 
In combination, scans with low ensemble uncertainty need less annotator time while yielding similar future DSC improvements. They are thus ideal candidates to enlarge a training set for upper-body distinct bone segmentation from CT scans.
}

\keywords{Distinct bone segmentation, Deep learning, CT, Active learning}

\maketitle

\section{Introduction}\label{sec1}

Automated segmentation of distinct bones within upper body CT scans opens up a world of possibilities. It supports surgical planning and navigation by providing semantic information to these systems.  For computations of joint load through bone density \cite{mullergerbl1989computed}, it simplifies preprocessing by separating adjacent bones. 
Furthermore, bone segmentation is a prerequisite for many types of diagnostics and analysis \cite{lindgren_belal_deep_2019}.

Bones are well visible in CT scans thanks to their characteristically high Hounsfield unit (HU) values. 
Semi-automated segmentation approaches usually consist of thresholding steps or region-growing from seeds, followed by manual tidying up of the edges and removing outliers and holes \cite{arguello2019comparison}. While they offer much control over the outcome, the necessary intermediate manual steps can be lengthy and cumbersome.

In recent years, neural networks have proven themselves to be handy tools for various semantic segmentation tasks involving medical images \cite{klein2019automatic,ronneberger2015unet,isensee2021nnu,gatti2021automatic,wolleb2020descargan,horvath2018spinal}.
Convolutional neural networks are also popular choices leading to good results for fully automated bone segmentation in full-body CT images and compare favourably to thresholding approaches \cite{klein2019automatic,leydon2021bone,noguchi2020bone}. However, these works focus on a binary segmentation task, separating bone tissues from the background without distinguishing single bones.

Multi-label segmentation of distinct bones poses additional challenges since it requires a very accurate separation of joint surfaces.
Atlas segmentation and explicit joint modelling are used in \cite{fu_hierarchical_2017}, who show excellent results for the segmentation of 62 distinct bones at the expense of numerous processing steps and a long inference time.
Most of the recent publications, however, steer more towards deep learning methods: 
In \cite{boutillon2020multi} five distinct bones are segmented using shape prior regularisation in combination with adversarial networks.
Their study also compares individual convolutional networks trained on one bone each, to those that segment multiple bones at once and shows that the latter lead to higher performing networks.
At large, most deep-learning-based medical imaging segmentation tasks deal with a relatively moderate amount of distinct bones. 
A rare exception is \cite{lindgren_belal_deep_2019} which uses a segmentation network in conjunction with a shape model-based landmark detection to segment and differentiate 49 bones. 
Since a priori it is not clear, whether an end-to-end learnt segmentation approach with more than twice the labels would be possible, we studied the task in \cite{schnider20203d} to understand the influence of a high number of labels on possible network architectures. 
In accordance with \cite{isensee2021nnu} we found that lean U-Net-like networks were the most suitable for distinct bone segmentation. 

Fully supervised semantic segmentation results improve with the dataset size \cite{mahapatra2018efficient}. 
In cases where large open datasets are missing -- such as for our task of distinct bone segmentation in upper bodies -- ground truth data is generally scarce and expensive to collect. 
Collecting and labelling a new dataset is always a challenge, but even more so in medical 3D segmentation, where obtaining ground truth data is a highly time-consuming task that needs to be carried out by specialists. 
In our case of distinct bone segmentation with many distinct labels, it takes multiple working days to segment a single CT scan from scratch. 
In that respect, a suitable strategy to automatically pre-segment scans, which then only need to be manually corrected, can save precious time. 

Methods to minimise annotator time are investigated under the term active learning \cite{budd2021survey}. 
Uncertainties within a network or between multiple networks can serve as an estimate of an unlabelled scan's future usefulness within a training dataset.
Special 2D segmentation networks have been proposed to be used together with bootstrapping in active learning \cite{yang2017suggestive}.
To avoid the costly retraining of the same model, Monte Carlo drop-out sampling has been used \cite{gal2017deep} and more elaborate metrics have been combined with it to estimate the representativeness of to-be-annotated data  \cite{ozdemir2018active}.  
As a draw-back, these approaches need a particular network architecture and require either frequent retraining of the same model or the presence of drop-out layers. In terms of 3D segmentation, this leads to additional challenges because the range of computationally feasible networks is much smaller than for 2D cases, where most active learning research has been conducted \cite{sourati2018active,mahapatra2018efficient}. Furthermore, many of the most successful 3D segmentation networks do not include drop-out layers, \cite{isensee2021nnu,cicek20163dunet} and their training is a very time-consuming affair. Alternatives that work without the need for drop-out layers, or a specific model architecture, are test-time augmentation \cite{wang2019aleatoric} and ensemble-based uncertainties \cite{beluch2018power}. 


In the following, we show how the combination of test-time data augmentation and model ensembles leads to robust results for distinct bone segmentation using as few as three scans in the training data. Furthermore, we provide an uncertainty measure based on test-time augmentation that works on any network architecture. Due to its plug-and-play nature, the method works on its own when time or space restraints hinder more complex means or when a network is delivered as-is. The uncertainty serves as an estimator for the number of voxels that need correction after the automated segmentation and can serve as a proxy to choose the least time-intensive new scans to label and include into the training data.

\section{Methods}


\subsection{Network design and training}
In previous experiments, we found 2D networks to work substantially worse for the task of distinct bone segmentation than 3D networks \cite{schnider20203d}.  
As a result, we concentrate on 3D networks in this current work. Unfortunately, this restricts our architecture choices due to the 3D networks' high demands in computational time and memory. This issue is exacerbated further by our task's high number of labels \cite{schnider20203d}.
For our experiments, we therefore choose a lean variation of the  3D-Unet architecture \cite{isensee2021nnu}. Thanks to its linear upsampling in place of up-convolutions, this architecture has a reduced number of trainable parameters than comparable networks \cite{cicek20163dunet}. With its resulting small memory footprint and the capability to adapt to a big range of segmentation tasks, it is well suited for our task.

We use the implementation provided within NiftyNet 0.5.0 \cite{gibson2018niftynet} for training and inference of our models. Due to the high computational demands segmenting numerous classes at once, we use a batch size of 1.
The model uses instance normalisation but no drop-out layers, which excludes the possibility of using Monte Carlo drop-out sampling. We run 20'000 training iteration steps per model, using the Adam optimiser with an initial learning rate of 0.001. For the training, we use Quadro RTX 6000 (24GB) and A100 SXM4 (40GB) GPUs. The training of a model takes 48 hours on the Quadro and 24 hours on the A100.

In terms of training time data augmentation, the patch-wise approach of training a 3D network already serves as a random cropping augmentation step.
Further, we apply randomised affine transformations, such as rotations upon the data before using it for training.

\subsection{Uncertainty estimation}\label{uncertainty}
The segmentation uncertainty of a non-labelled image can be estimated using multiple predictions from learnt models \cite{ozdemir2021active}.
The uncertainty in classification $y_v$ of a single voxel $v$ as belonging to class $l = 1,...,L$ by $N$ different predictions with input $x$ and model parameters $\theta$ is given as:
\begin{equation}
    \mathrm{uc}_{l,y_{v}}=\mathrm{var}\left[p(y_{v}=l\vert x,\theta) \right]\, ,
\end{equation}
 where $p(y_{v}=l \vert x,\theta)$ is the vector containing the probability of voxel $y_v$ being classified as $l$ for all $N$ predictors. Since all voxels and all labels are equally important, we compute the unweighted average over the volume and labels:
\begin{equation}\label{uncertainty_equation}
    \mathrm{uc}=\frac{1}{L}\sum_{l=1}^{L}\frac{1}{V}\sum_{v=1}^{V}\mathrm{var}\left[p(y_{v}=l \vert x,\theta) \right] \, .
\end{equation}

The multiple predictions used to compute the uncertainty can be obtained in several ways, which can be used independently or in combination. 
In the ensemble approach, various models provide one prediction each. 

Conversely, test-time augmentation works on a single trained model and obtains multiple predictions by varying the treatment of the input data during inference. 
Copies of the input data are transformed, the inference is performed, and the resulting predictions are then transformed back into the original space.
Hence, only invertible types of data augmentation are suitable.
Affine transformations can be inverted but generally suffer from errors due to the necessary interpolation into pixel space, which will occur twice. 
In contrast, offsets or translations by an integer number of pixels can be inverted without introducing new error sources, leading to different inference results.

Apart from their use as uncertainty estimators, predictions of the same input can be combined with a voting scheme, such as majority voting, to create an ensemble prediction \cite{breiman1996bagging,dietterich2000ensemble,feng2020brain,isensee2021nnu}. We will use these multiple predictions of the same input for both, the uncertainty estimate, and to create ensembles.

\section{Experiments}
We test the segmentation capabilities of our ensemble approach for various types of bones. We also investigate the ensemble uncertainty of scans that could potentially enlarge the dataset in relation to the gained segmentation accuracy and the manual correction time needed.

\subsection{Datasets}

\begin{table}[h]
\caption{Dataset properties.}\label{datatab}
\begin{tabular}{l l l l l l }

\toprule
\multicolumn{1}{l}{Dataset}    & 
\multicolumn{1}{l}{volumes} & 
\multicolumn{1}{l}{male/female} &
\multicolumn{1}{l}{age}  &
\multicolumn{2}{l}{original resolution (mm)}\\
\cmidrule{5-6}


\multicolumn{4}{c}{\phantom{i}}& 
\multicolumn{1}{l}{in-plane} &
\multicolumn{1}{l}{out-plane}\\

\midrule

A, inital& 5 & 2/3 & 44-60 & 0.83-0.97 & 1.0-1.5 \\ 
B, follow-up& 12 & 7/4 &54-103 & 0.89-0.98 & 1.0-1.3\\ 
\addlinespace

\bottomrule
\end{tabular}
\end{table}










The 16 CT scans we use (see Table \ref{datatab}) were routinely obtained post mortem from body donors at our university's anatomical department. 
Due to limited annotation resources, we had to choose a small fraction of all available data for manual segmentation. We excluded scans with rare skeletal variants, as well as scans containing implants that led to strong artefacts.  
All scans were taken using the same scanner and  field of view, which starts at the top of the skull and stops approximately mid-femur.
All body donors lie on their backs, arms either resting on the lap or crossed over the stomach to various degrees.  Including background, 126 classes have been segmented, spanning all upper body bones, pelvis, and femurs. 
the subjects' advanced age (see Table \ref{datatab}) manifests in different levels of scoliosis and calcifications.



Dataset A has already been used and described in \cite{schnider20203d}.
The remaining eleven scans of dataset B were pre-segmented using our ensemble models, and the labelling afterwards manually improved. Nevertheless, the time needed to finish a manual segmentation still exceeded a working day per scan due to the many distinct classes.
We re-sample our data to a uniform resolution of $\SI{2}{\milli\metre}$ in all three dimensions which leaves us with scans of $\sim 265 \times 256 \times 512$ voxels. 
The smaller resolution reduces I/O times during training tremendously and allows us to capture more body context within an input patch.
The labels of the various bones are highly imbalanced, which complicates model training. While many bones in the hands only span a few dozen voxels, there are also big bones such as the skull, femur, and pelvis, which easily exceed 10'000 voxels. 

\begin{figure*}[t]
\includegraphics[width=\textwidth]{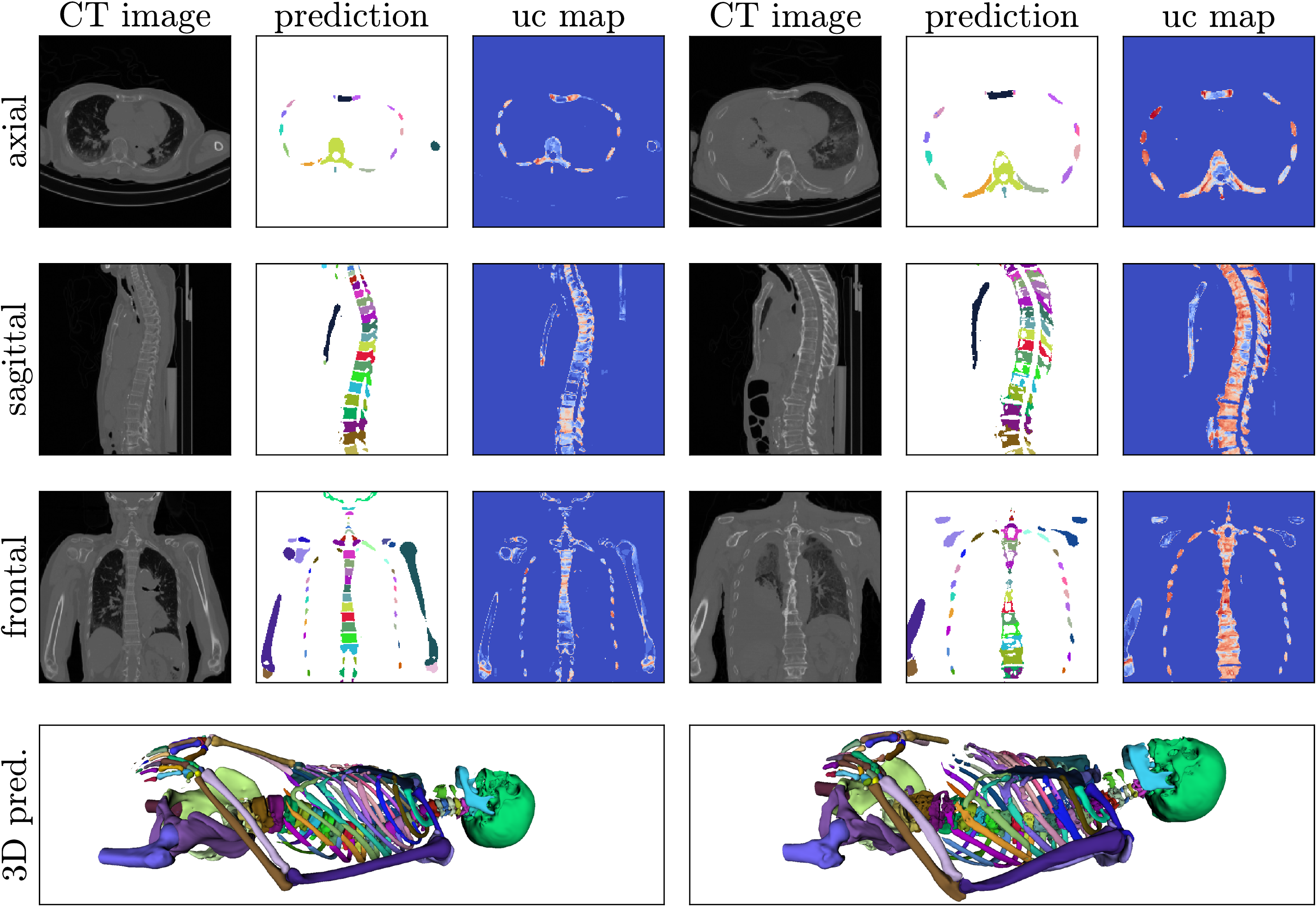}
\caption{Qualitative results were obtained using the ensemble approach of combining six model variations on a hold-out set of unlabelled scans. Shown are CT images predicted label map, and uncertainty map for both the scan with lowest (left) and highest (right) mean uncertainty \eqref{uncertainty}. 
} \label{results}
\end{figure*}



\subsection{Ensemble Model Variations}
To obtain different predictors for our ensemble, we perform the training using a range hyperparameter settings that have been found to work well for the given task \cite{schnider20203d}. 
Input patch sizes range from \SI{96}{\px} - \SI{160}{\px} per dimension, the loss function used is either pure multi-label cross-entropy ({\small\textbf{xent}}) or a linear combination of the DSC loss \cite{milletari2016v} and the cross-entropy loss ({\small\textbf{D+xent}}). Patches are sampled uniformly random ({\small\textbf{unif}}) or in a balanced fashion ({\small\textbf{bal}}), such that the probability to be present in the patch is the same for all labels.
We use elastic deformations ({\small\textbf{d}}) and slight affine transformations ({\small\textbf{a}}) for data augmentation.
The six variations used for the experiments are 
$160_{\textrm{bal,D+xent,d}}$, $160_{\textrm{bal,xent,d}}$, 
$128_{\textrm{unif,D+xent,a,d}}$, 
$128_{\textrm{unif,D+xent,d}}$,
$128_{\textrm{bal,D+xent}}$,
$128_{\textrm{bal,xent}}$, where the numbers and indexes represent the modes explained above.

\subsection{Evaluation metrics}\label{metrics}
To measure the segmentation performance per class $c$ we use the Dice similarity coefficient: $\mathrm{DSC}_c = \frac{2\abs{P_c\odot G_c}}{\abs{P_c}+\abs{G_c}}$, with $P_c$ the pixel-wise prediction of class $c$, and $G_c$ the corresponding ground truth. To indicate performance over a group of bones, we give the median, as well as the 16 and 84 percentile of the corresponding $\mathrm{DSC}_c$ scores. Classes which were missed completely by the prediction are not included to not distort the distribution, we give the detection ratio $\mathrm{dr}\coloneqq \frac{\#\text{ bones with DSC}>0}{\#\text{ all bones}}$ to account for them.

We use 5-fold cross-validation. Within each fold of dataset A, three scans are assigned to training, one to validation, and the remaining scan serves as the test set. For every cross-validation fold, the test set is different.
When training in conjunction with dataset B, we keep the test set per fold consistent to facilitate the comparison of results. The new data B are added to the training (10) and validation (2) splits.

\section{Results}
 
Our initial experiments showcase the astonishingly good results that can be achieved for a tiny dataset on the challenging task of distinct bone segmentation, segmenting 126 classes simultaneously and end-to-end.
Quantitative results are given in Table \ref{tab1}. We compare the average performance of our single models (\textbf{Sngl $\varnothing$}) with different types of ensembles. As a baseline we also provide results for $160_{\textrm{bal,D+xent,d}}$, the best performing single model (\textbf{Bsm}).

\begin{table}
\caption{Ablation results using single models and various ensemble types on dataset A. For comparison we also give results on the final full dataset A+B.}\label{tab1}
\begin{tabular}{l r r r r r r r}

\toprule



& \multicolumn{1}{c}{Vertebrae} & \multicolumn{1}{c}{Ribs} & \multicolumn{1}{c}{Hands} & \multicolumn{1}{c}{Large} & \multicolumn{1}{c}{All}&es\footnotemark[1]&ds\footnotemark[2]\\

\midrule

Sngl$\varnothing$&
$0.81_{-0.24}^{+0.09}$& 
$0.58_{-0.28}^{+0.22}$&
$0.54_{-0.41}^{+0.26} (52\%)$& 
$0.88_{-0.10}^{+0.05}$&
$0.72_{-0.40}^{+0.17} (78\%)$& 
1&3\\ 
\addlinespace
Bsm&$0.82_{-0.17}^{+0.09}$& 
$ 0.58_{-0.27}^{+0.21}$&
$ 0.67_{-0.39}^{+0.18} (58\%)$&  
$ 0.88_{-0.11}^{+0.04}$&
$ 0.75_{-0.38}^{+0.14} (80\%)$& 
1 &3\\ 
\addlinespace

\midrule

$\mathrm{Bsm_{so}}$&
$0.86_{-0.19}^{+0.06}$& 
$0.65_{-0.33}^{+0.19}$&
$0.58_{-0.38}^{+0.24}(58\%)$& 
$0.89_{-0.07}^{+0.04}$&
$0.76_{-0.38}^{+0.14} (81\%)$& 
7 &3\\ 
\addlinespace
$\mathrm{Bsm_{a+so}}$&
$0.85_{-0.17}^{+0.07}$&
$0.60_{-0.31}^{+0.21}$&
$0.56_{-0.39}^{+0.22}(54\%)$& 
$0.89_{-0.08}^{+0.04}$&
$0.75_{-0.40}^{+0.14} (79\%)$&  
7&3 \\ 
\addlinespace
$\mathrm{Ens}$& 
$0.87_{-0.18}^{+0.06}$ & 
$0.66_{-0.33}^{+0.20}$ &
$0.65_{-0.50}^{+0.19} (57\%)$  & 
$\mathbf{ 0.93_{-0.03}^{+0.03}}$ &
$0.80_{-0.39}^{+0.12} (80\%)$  & 
6 &3\\ 
\addlinespace
$\mathrm{Ens_{so}}$& 
$0.88_{-0.18}^{+0.06}$ & 
$0.68_{-0.31}^{+0.19}$ &
$0.62_{-0.48}^{+0.24}(52\%)$  & 
$0.92_{-0.03}^{+0.03}$ &
$0.83_{-0.41}^{+0.10} (78\%)$  & 
42 &3\\ 
\addlinespace
\midrule
 A+B &
$\mathbf{0.89_{-0.15}^{+0.04}}$ & 
$\mathbf{0.80_{-0.15}^{+0.07}}$ &
$\mathbf{0.79_{-0.37}^{+0.09}(69\%)}$  & 
$0.91_{-0.08}^{+0.04}$ &
$\mathbf{0.83_{-0.22}^{+0.08} (85\%)}$  & 
1&13 \\ 
\addlinespace
\addlinespace
\botrule
\end{tabular}
\footnotetext[]{Results in DSC with the detection rate dr in brackets if it is less than 100\%.}
\footnotetext[1]{The number of inferences used to compute the results is indicated as ensemble size es.}
\footnotetext[2]{The number of images in the training split is indicated as dataset size ds.}
\end{table}

\begin{figure*}[t]
\includegraphics[width=\textwidth]{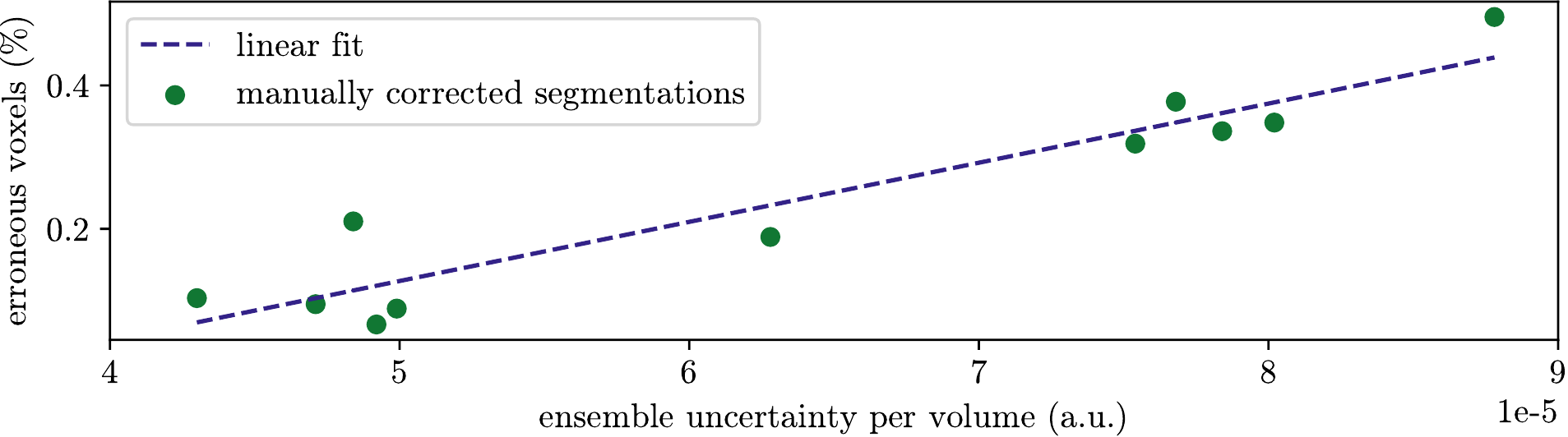}
\caption{A higher ensemble uncertainty leads to worse priors requiring more manual work to correct them. While the percentages seem minor, the absolute values of voxels that need correction span from 20767 to 115291.} \label{uncertainty_error_corr}
\end{figure*}

\begin{figure*}[t]
\includegraphics[width=\textwidth]{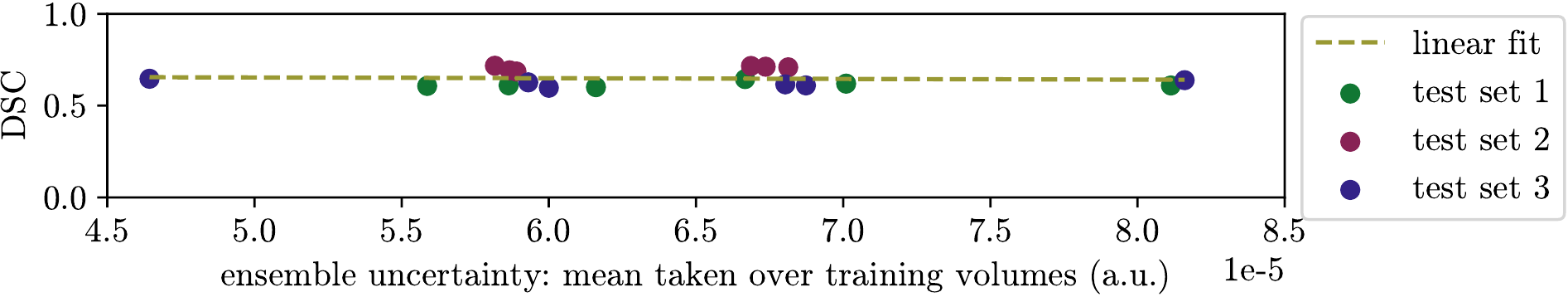}
\caption{Comparison of summed mean uncertainty of the training set and the resulting Dice score. We use three different test sets combined with 6 training sets each.} \label{uncertainty_DSC}
\end{figure*}

We compare ensembles built using differently trained models (\textbf{Ens}), and the best single model using test-time data augmentation in the shape of sampling offset ({\small\textbf{so}}), and affine transformations ({\small\textbf{a}}). As expected, the more variants of models and test-time augmentation we use, the better the final results. A depiction of qualitative segmentation results and uncertainty maps can be found in Figure \ref{results}.

For the follow-up experiments, we created a ground truth labelling for dataset B by using the hitherto best ensemble prediction $\mathrm{Ens_{so}}$ to create a segmentation that was then cleaned up manually.
We provide qualitative segmentation results achieved when training the initial dataset A alongside dataset B as a comparison to our ensemble results. Using more data -- training with 13 instead of 3 scans -- naturally improves results. The improvement is particularly evident for hands and ribs. On vertebrae and large bones, the difference is much smaller and the ensemble performs almost as good as the model trained on both A and B.

To analyze the use of the uncertainty metric, we defined three test sets, and for each of those then six training sets consisting of dataset A plus three scans from dataset B. The uncertainty of the scans in B had been established using the ensemble $\mathrm{Ens_{so}}$. 
We trained the resulting 18 cases and plot the mean uncertainty of the training set against the DSC achieved on the test set  (see Figure \ref{uncertainty_DSC}). The influence of the uncertainty on the segmentation performance is surprisingly small.

In Figure \ref{uncertainty_error_corr}, we plot the mean uncertainty derived from $\mathrm{Ens_{so}}$ against the volume-normalized percentage of voxels that needed to be corrected during the ground truth segmentation of dataset B. If we take the number of voxels that need correction as a surrogate for the manual effort needed, a higher uncertainty leads to up to 5 times more correction effort.


\section{Conclusion}
In this work, we examined the correlation between ensemble uncertainty and the number of erroneous voxels the ensemble produces for the task of distinct bone segmentation. 
On the one hand, we found a correlation between uncertainty and erroneous voxels, implying that scans with low ensemble uncertainty tend to be more accurately segmented. On the other hand, the correlation between a scan's uncertainty and the prospective DSC change caused by incorporating said scan into the training set was negligible.

As a result, when planning to increase the size of the available dataset, the uncertainty  measure can be used to choose so-far unlabelled scans to minimise the time for new annotations. 
Since low-uncertainty scans need the least of the annotators' time, while leading to the same improvement of DSC, their choice maximises the total amount of newly labelled scans in a given time budget.

We also explored the use of test-time data augmentation as part of an ensemble method for distinct bone segmentation, where only very little labelled data is available. We observed that the ensemble approach achieves the same performance as does training on an enlarged training set of three times as many scans.







\backmatter



\bmhead{Acknowledgments}

This work was financially supported by the Werner Siemens Foundation through the MIRACLE project.

\section*{Declarations}


\begin{itemize}


\item Competing interests:
None of the authors have competing interests to declare that are relevant to the content of this article.
\item Funding:
This work was financially supported by the Werner Siemens
Foundation through the MIRACLE project.
\item Ethics approval:
This research study was conducted retrospectively from CT data routinely obtained from body donours. Thus no ethical approval is required.
\item Consent to participate:
Informed consent was obtained from all individual body donors included in the study.
\item Consent for publication:
Body donors signed informed consent regarding publications using their data.
\item Availability of data and materials:
 The dataset is not publicly available.
\item Code availability:
The code for the segmentation and uncertainty computations can be shared on request.

\end{itemize}

\noindent






\begin{appendices}






\end{appendices}


\bibliography{sn-bibliography}


\end{document}